\documentclass[
    a4paper,
    man,
    british,
    floatsintext
]{apa6}

\usepackage[british]{babel}
\usepackage{amsmath}
\usepackage[utf8]{inputenc}
\usepackage{epstopdf}
\usepackage{csquotes}
\usepackage{color}
\usepackage{xcolor}
\usepackage{changes}
\usepackage[hidelinks]{hyperref}
\usepackage{caption}
\usepackage{subcaption}
\usepackage{epstopdf}
\usepackage[
    style=apa,
    backend=biber,
    sortcites=true,
    sorting=nyt,
    hyperref=false,
    backref=false
]{biblatex}

\makeatletter

\renewcommand{\@seccntformat}[1]{\csname the#1\endcsname.\quad}
\renewcommand\section{\@startsection{section}{1}{\z@}%
	{-3.5ex \@plus -1ex \@minus -.2ex}{2.3ex \@plus .2ex}{\normalfont\large\bfseries}}
\renewcommand\subsection{\@startsection{subsection}{2}{\z@}%
	{-3.25ex \@plus -1ex \@minus -.2ex}{1.5ex \@plus .2ex}{\normalfont\normalsize\bfseries}}
\makeatother

\DeclareLanguageMapping{british}{british-apa}

\let \cite \parencite

\DeclareLanguageMapping{british}{british-apa}
\title{Exploring the Dynamics of External and Self-Citations and Their Role in Shaping Scientific Impact}
\shorttitle{Exploring the Dynamics of External and Self-Citations and Their Role in Shaping Scientific Impact}
\author{
    Maciej J. Mrowinski$^{1,*}$,
    Aleksandra Buczek$^{1}$,
    Agata Fronczak$^{1}$    
}

\affiliation{
    $^{1}$ Warsaw University of Technology, Faculty of Physics,\\
    ul.~Koszykowa 75, 00-662 Warsaw, Poland\\
    $^{*}$ Corresponding author; email: maciej.mrowinski@pw.edu.pl
}

\usepackage{etoolbox}
\makeatletter
\preto\maketitle{\setcounter{secnumdepth}{-1}}
\appto\maketitle{\setcounter{secnumdepth}{3}}

\patchcmd{\maketitle}{\section{\@title}}{}{}{}   
\makeatother

\abstract{Understanding the mechanisms driving the distribution of scientific citations is a key challenge in assessing the scientific impact of authors. We investigate the influence of the preferential attachment rule (PAR) in this process by analysing individual citation events from the DBLP dataset and two Scopus-based datasets, enabling us to estimate the probability of citations being assigned preferentially. Our findings reveal that, for the aggregated dataset, PAR dominates the citation distribution process, with approximately 70\% of citations adhering to this mechanism. However, analysis at the individual level shows significant variability, with some authors experiencing a greater prevalence of preferential citations, particularly in the context of external citations. In contrast, self-citations exhibit notably different behaviour, with only 20\% following PAR. We also demonstrate that the prominence of PAR increases with an author’s citability (average citations per paper), suggesting that more citable authors are preferentially cited, while less-cited authors experience more random citation patterns. Furthermore, we show that self-citations may influence bibliometric indices, such as the $h$-index. Our results confirm the distinct dynamics of self-citations compared to external citations, raising questions about the mechanisms driving self-citation patterns. These findings provide new insights into citation behaviours and highlight the limitations of existing approaches.}

\begin{document}
\maketitle

\setcounter{section}{0}
\setcounter{secnumdepth}{3}
\section{Introduction}

Numerous models have been proposed to describe how citations are distributed. These models operate either at the citation network level (see, e.g., \textcite{peterson2010}) or at the author level (see \textcite{Siudem2020}), where the aim is to reconstruct an author’s citation vector from selected bibliometric parameters. Here we focus on the latter. A citation vector records an author’s publication and citation history; each element corresponds to the number of citations received by the corresponding paper in order of publication. For example, \textcite{Ionescu2013} introduced a model that distinguishes between self-citations and external citations (i.e., all citations not classified as self-citations).

It is worth mentioning that the definition of self-citations is not straightforward, as they can be defined and counted in various ways \cite{Ioannidis2015}, which may lead to confusion \cite{Li2020}. In its simplest (direct) form, a self-citation is counted as any citation by an author to one of their own previous papers. Additionally, the definition can be extended to include citations made by coauthors of a paper, often referred to as co-author self-citations. The most challenging - and in some cases nearly impossible - type of self-citations to identify are coerced self-citations \cite{Thombs2015}. These occur indirectly, such as when reviewers request that authors include references to specific articles during the peer review process. In the results presented in this paper, we focus exclusively on direct self-citations.

The Ionescu-Chopard model assumes that both types of citations are distributed according to the preferential attachment rule (PAR). PAR, often referred to as the "rich get richer" principle \cite{Perc2014, Price1963-book}, posits that articles with many citations are more likely to attract additional citations. However, intuition and studies suggest that self-citations may not follow this rule. For example, authors might be more likely to self-cite their most recent articles \cite{Shah2015, Aksnes2003}. It is also worth noting that the Ionescu-Chopard model does not explicitly describe the dynamics of citation distribution. Instead, it serves as an artificial procedure designed to reproduce the citation vector at a specific point in time.

Like the Ionescu-Chopard model, many citation models assume some form of preferential attachment. Intuitively, this seems reasonable. When compiling bibliographies or searching for relevant literature, authors often use scientific databases, likely favouring highly cited papers over less cited ones. Moreover, the citation network is scale-free \cite{Redner1998} (the variance of the node degree is effectively infinite, rendering the average node degree meaningless), a property frequently attributed to preferential attachment. Nevertheless, this raises a critical question: how significant is PAR in the citation distribution process? In this paper, we aim to determine the true fraction of preferential citations and address this question.

Another key question concerns self-citations, a topic that is both fascinating and contentious in the scientific community. Some view self-citations as a hallmark of productive authors \cite{Mishra2018}, arguing that they can genuinely enhance the visibility of one's work \cite{GonzalezSala2019}, which in turn boosts external citations with little downside \cite{Fowler2007}. However, while there are many legitimate reasons for self-citation \cite{Brysbaert2011, Pichappan2002}, some are less benign. Self-citations can be exploited to artificially inflate bibliometric indicators \cite{Loan2021, Amjad2020}, a practice that should be considered when evaluating scientific impact \cite{Davarpanah2009}. Policymakers must also exercise caution when designing systems for career advancement and evaluation, as such policies can unintentionally influence self-citation behaviour \cite{Abramo2021, Peroni2020, Seeber2019}. Interestingly, self-citation issues extend beyond individual authors and can affect entire journals \cite{Taskin2021, Hartley2012}, where they may be used to manipulate journal impact factors. Given their importance, it is crucial to understand the properties of self-citation distribution. Is it similar to external citation distribution? Does preferential attachment play a key role? These are questions we will address in this paper.

To determine the fraction $\rho$ of citations governed by the preferential attachment rule and to answer the questions posed above, we adopted a largely data-driven approach. Using the extensive DBLP database and two Scopus-based datasets, we reconstructed citation histories and vectors for a large set of authors, tracking the evolution of citations over time. By adapting a methodology previously employed to study preferential attachment in complex networks \cite{Leskovec2008}, we quantified the fraction $\rho$ of citations governed by PAR.

\begin{figure}[!t]
    \centering
    \includegraphics[width=0.5\linewidth]{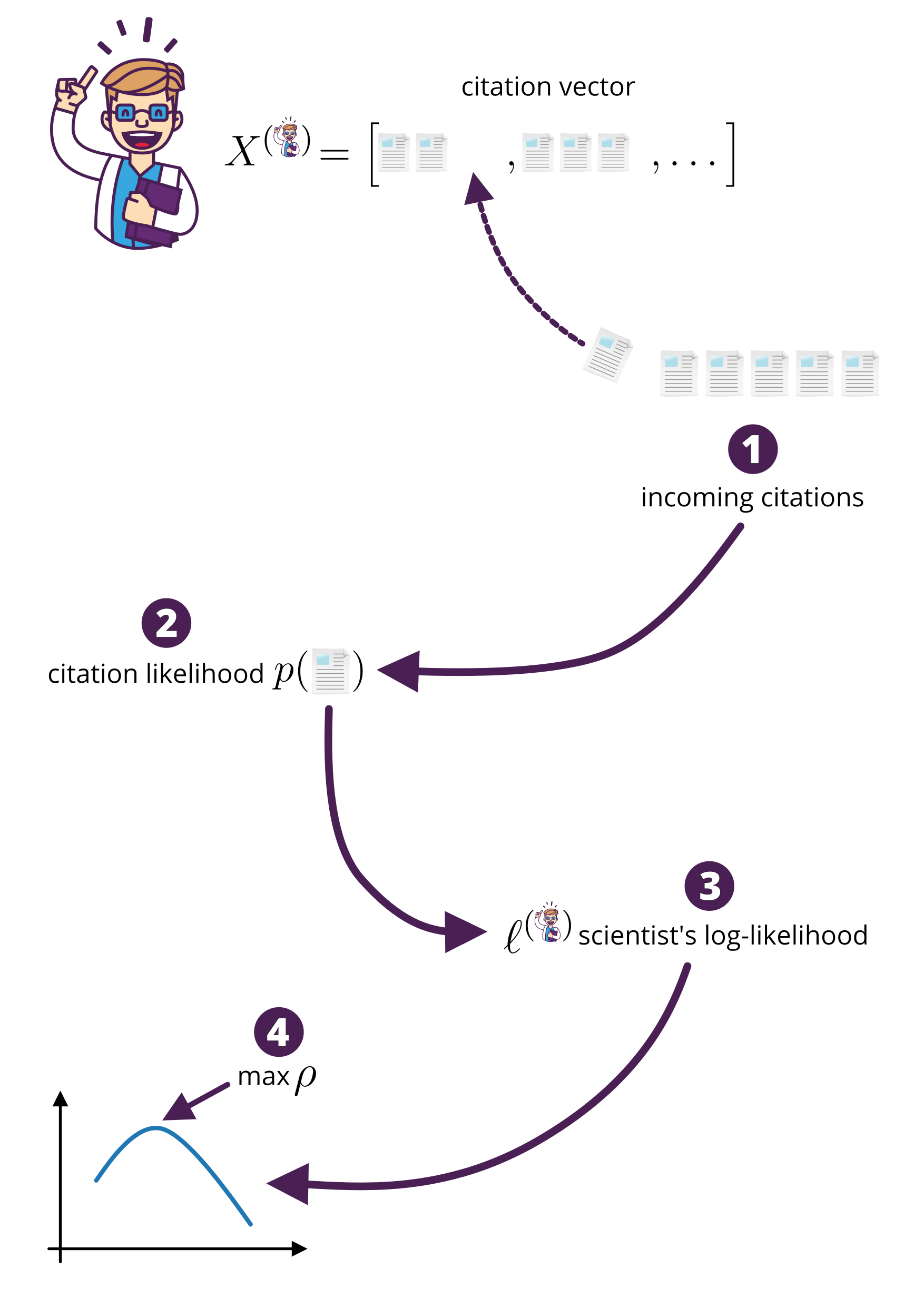}
    \caption{The process of calculating the value of $\rho$ for a specific scientist involves the following steps: 1) Each citation is processed individually, in the order it was received. 2) The probability of each citation is computed based on the assumptions of the model. 3) These probabilities are combined to form the log-likelihood. 4) The log-likelihood is maximised, resulting in the value of $\rho$.}\label{fig:ll-calculation}
\end{figure}
Our approach involves calculating the likelihood of each citation event based on the state of an author’s citation vector at the previous time step, incorporating both preferential attachment and random allocation mechanisms. The parameter $\rho$, which controls the balance between these mechanisms, is estimated by maximising the log-likelihood for individual authors (as illustrated schematically in Figure~\ref{fig:ll-calculation}) and for the dataset as a whole. This analysis was conducted for all types of citations combined, as well as separately for self-citations and external citations, providing a detailed characterisation of citation dynamics.

\section{Methodology}
\subsection{Model}

One of the most significant and profound discoveries in the study of complex networks was the realisation that many of their properties can be explained by the preferential attachment rule \cite{albert2002}. However, while it is possible to construct models that replicate certain characteristics of real systems, this does not necessarily mean that the dynamics and assumptions of these models align with those of the actual systems they aim to describe.

In \textcite{Leskovec2008}, the authors employed a simple yet ingenious method to investigate preferential attachment in networks. They analysed the temporal evolution of a real network, link by link, and calculated the probability of each link forming based on the assumptions of several proposed models. This approach allowed them to aggregate the probabilities of the network's formation under different models and to compare the likelihood of each model. We have adapted this methodology to examine the role of the preferential attachment rule in the process of citation distribution. This subsection provides a detailed description of the procedure we employed.

Scientist $j$ is characterised by a citation vector
\begin{equation}
    X^{(j)}(t)=\left[X^{(j)}_1(t), X^{(j)}_2(t),\dots\right],
\end{equation}
where $X^{(j)}_i(t)$ denotes the number of citations of the $i$-th article at the $t$-th time step. We assume that time is discrete and that the length of the citation vector $N^{(j)}(t)=\left|X^{(j)}(t)\right|$ can grow with time as new articles are published. Due to the granularity of the available data, one time step corresponds to one year.

The author receives a number of citations during each time step. These citations can also be arranged in a vector:
\begin{equation}
    C^{(j)}(t)=\left[C^{(j)}_1(t), C^{(j)}_2(t),\dots\right],
\end{equation}
where $C^{(j)}_i(t)$ is the index in the citation vector of the article that received the citation.

If we define a model $p(k|X,\rho)$ that specifies, given the citation vector $X$ and some parameter $\rho$, the probability of the $k$-th article receiving a citation, we can calculate the log-likelihood for this model (for a single author) in the following way:
\begin{equation}
    C^{(j)}(\rho) = \sum_t\sum_{k \in C^{(j)}(t)} \ln p(k|X^{(j)}(t-1),\rho). \label{eq:logl}
\end{equation}
The first summation in this equation is taken over the entire career of author $j$, or up to the most recent point for which data is available. This log-likelihood can then be numerically maximised, yielding the value of the parameter $\rho$ that best describes the citation patterns of author $j$.

It is important to note that the distribution of citations at the $t$-th time step depends on the state of the citation vector from the previous time step, $t-1$. However, articles published during the $t$-th time step are added to this vector with their citation count initialised to zero. This approach accounts for the fact that we can only determine the year of publication, and some articles receive citations within the same year they are published. In fact, due to the peculiarities of online publishing, some articles even receive citations before their publication date, but we disregard these cases.

The log-likelihood for the entire dataset can be calculated as the sum of log-likelihoods for all authors:
\begin{equation}
    \ell(\rho) = \sum_j \ell^{(j)}(\rho). \label{eq:loglall}
\end{equation}
This value can also be maximised, yielding $\rho$ characterising the combined dataset.

The model we study in this paper is a mixture of pure randomness (uniform distribution) and preferential attachment
\begin{equation}
    p(k|X(t-1),\rho) = \rho \frac{X_{k}(t-1)}{\sum_i X_i (t-1)}+(1-\rho)\frac{1}{N(t)}.\label{eq:model}
\end{equation}
We emphasise one more time that while the model depends on the state of the citation vector from the previous time step, we add to it articles published in the current time step with their number of citations initialised to $0$ - hence $N(t)$ instead of $N(t-1)$ in the random term. Also, we do not artificially add $1$ to the number of citations in order to kickstart the preferential attachment mechanism. What follows is that the first citation of an article must come from the random term. We believe it is a reasonable assumption, but it also means a pure preferential model with $\rho=1$ is not valid - the probability of each first citation, and consequently the entire likelihood, would be $0$.

Finally, we must address an important issue related to the interpretation of the parameter $\rho$. While it might be tempting to view $\rho$ as a parameter ranging from pure randomness (for $\rho=0$) to pure preferential attachment (for $\rho=1$), this interpretation oversimplifies the situation. Although a random component is undoubtedly present in the citation distribution process, many additional factors are likely at play, and a more complex model than the one in Eq.~(\ref{eq:model}) could be constructed. Therefore, it is more prudent to interpret the scale of $\rho$ as spanning the range between non-PAR and PAR, with the random term in the model functioning as a placeholder to account for citations that cannot be explained by PAR. This interpretation will be adopted throughout the paper.

\subsection{Dataset - DBLP}

We utilized the 12th version of the AMiner DBLP Citation Network Dataset for our study \cite{Tang2008}. This dataset is a comprehensive collection of metadata for approximately 4 million scientific articles, primarily in the field of computer science. The metadata includes information such as the year of publication, the list of authors, and the list of references, enabling the reconstruction of the citation network - a network where nodes represent articles and directed edges represent citations.

The citation network built from DBLP covers about 3 million authors; unless noted otherwise, our analyses focus on the $\sim 200,000$ with 10 or more publications (we do, however, use all of the available citation information for these authors - we do not filter out any publications from the database). In this subsample, 75\% of authors published their first article between 1998 and 2014, with the mode in 2006. For the papers in this subsample, 75\% were published between 2006 and 2019.

It is worth noting that, while extensive, the DBLP dataset represents only a sample of the complete citation network and we did not supplement the information in the dataset with external sources. It means that on average, other articles contained in DBLP cover about 60\% of all citations for each individual article (this number can be estimated by comparing the recreated network with the number of citations available in the metadata). Nevertheless, we believe this is sufficient for qualitative analysis, especially given the reasonable assumption that the missing citations are of a similar nature across all articles. Also, DBLP contains information on various types of publications - journal articles (approximately 36\%), conference proceedings (approximately 52\%), books, and others. We decided to include all of these types in the citation vectors. 

Two issues arise with datasets like DBLP. First, author disambiguation: ensuring that records with the same or similar names refer to different people. DBLP assigns each author a unique identifier; we relied on these IDs - rather than names - to distinguish authors and did not apply any extra disambiguation beyond DBLP’s own methods. Second, publication year reliability: a paper can have multiple dates (online, print, etc.), and the AMiner DBLP Citation schema does not specify which one is used. We found occasional cases where papers accumulated citations dated earlier than the cited paper’s year. Spot checks indicate this often stems from republication (e.g., conference proceedings later republished in a journal). We therefore excluded such events - fewer than 1\% per author - from our analysis.

\begin{figure}[!t]
    \centering
    \begin{subfigure}[t]{.48\textwidth}
    \captionsetup{position=top,justification=centering}    
    \centering
    \caption{}
    \includegraphics[width=\textwidth]{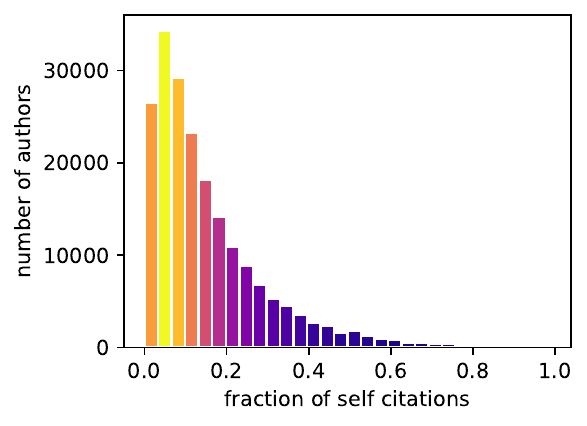}
    \end{subfigure}
    \begin{subfigure}[t]{.48\textwidth}
    \captionsetup{position=top,justification=centering}    
    \centering
    \caption{}    
    \includegraphics[width=\textwidth]{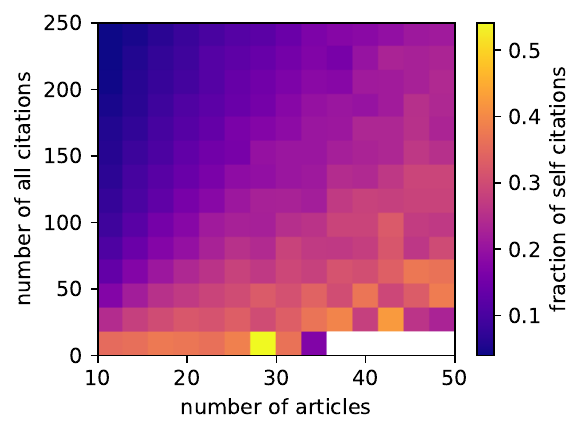}
    \end{subfigure}\\
    \begin{subfigure}[t]{.48\textwidth}
    \captionsetup{position=top,justification=centering}    
    \centering
    \caption{}
    \includegraphics[width=\textwidth]{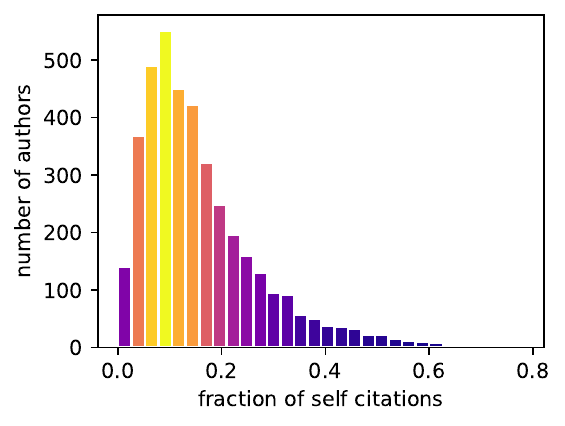}
    \end{subfigure}
    \begin{subfigure}[t]{.48\textwidth}
    \captionsetup{position=top,justification=centering}    
    \centering
    \caption{}
    \includegraphics[width=\textwidth]{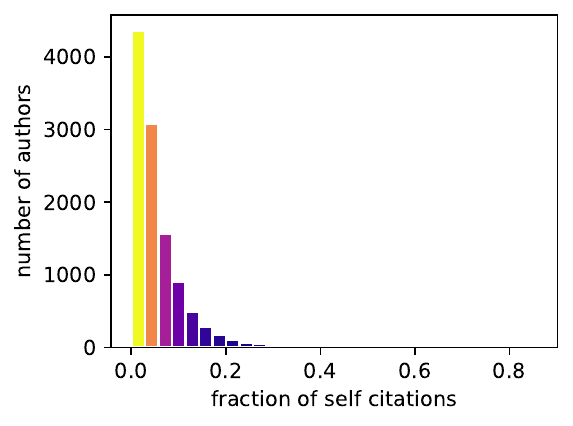}
    \end{subfigure}    
    \caption{Panel (a) presents a histogram of the fraction of self-citations (among all citations) for DBLP. Panel (b) shows the average fraction of self-citations for authors grouped by their total citation counts and number of published papers for DBLP. Panels (c) and (d) present a histogram of the fraction of self-citations for the PRE and Nature datasets, respectively.}\label{fig:selfcits}    
\end{figure}
Given that we will examine the distribution of self-citations, it is worthwhile to explore some characteristics of self-citation behaviour among authors in the DBLP dataset. After all, self-citations could represent only a small fraction of total citations and might be negligible, at least for this group of authors. Panel (a) of Figure~\ref{fig:selfcits} presents a histogram of the fraction of self-citations (the percentage of self-citations among all citations) for authors who have published 10 or more papers. The average fraction of self-citations is 16\%, and as shown, some authors exhibit significantly higher percentages of self-citations than the average. This indicates that self-citations are far from negligible.

This observation is corroborated by panel (b) of the same figure, which depicts the average fraction of self-citations for authors with varying numbers of published papers and total citations received. The data suggest that authors with fewer total citations relative to the number of published papers tend to have a higher fraction of self-citations. This result is intuitive, as external citations are expected to grow - and, except in extreme cases, should grow - at a higher rate over time. Additionally, this panel further highlights that self-citations are not negligible, as they constitute a significant proportion of total citations for many authors, particularly those with lower overall citation counts.

\subsection{Dataset - Scopus}
As mentioned earlier, while DBLP is a large dataset, it does not provide the complete citation network for the authors it includes. As a consequence, the citation vectors reconstructed using DBLP omit some of the authors’ publications, and the citation counts based on DBLP references are lower than their actual values. Additionally, DBLP is primarily focused on computer science.

To test the universality of our results, we repeated the analysis on two smaller but complete datasets. The first includes all authors who published an article in Nature in 2019, and the second includes authors who published an article in Physical Review E (PRE) in the same year. Both datasets were retrieved via the Scopus API and include full publication and citation histories. After removing authors with fewer than 10 papers or with no citations, the datasets contain about 11,000 authors for Nature and 4,000 for PRE, and our analysis is restricted to these authors.

In the Nature dataset, 75\% of authors published their first paper between 1993 and 2018; in the PRE dataset, between 1989 and 2017. The modes are 2010 and 2009, respectively. Regarding publications, 75\% fall between 2007 and 2020 in the Nature dataset and between 2007 and 2021 in the PRE dataset. Because annual output increases over time, most publications in both datasets lie between 2007 and the end of the download window (2020 to 2021).

Because Scopus is a curated database, these two datasets should have far fewer author-disambiguation problems than DBLP. For the same reason, publication dates should be more consistent. In both Scopus-based datasets, fewer than 0.1\% of citations per author predate the cited paper’s publication date. As with DBLP, we excluded these cases from the analysis.

A key difference between DBLP and the Scopus-based datasets is the venues. DBLP focuses on computer science, so most publications are conference proceedings. In the PRE dataset, however, the overwhelming majority (about 80\%) are journal articles, and conference proceedings make up about 14\%. The Nature dataset is similar, with 78\% journal articles and about 8\% conference proceedings. The average self-citation fraction is 16\% for PRE and 5\% for Nature; distributions are shown in Figure~\ref{fig:selfcits}.

\section{Results}

\subsection{Analysis of whole dataset}

\begin{figure}[!ht]
    \centering
    \begin{subfigure}[t]{.31\textwidth}
    \captionsetup{position=top,justification=centering}    
    \centering
    \caption{}
    \includegraphics[width=\textwidth]{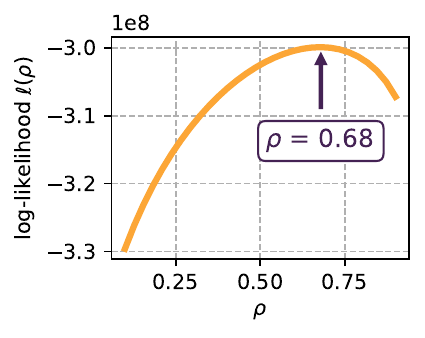}
    \end{subfigure}
    \begin{subfigure}[t]{.31\textwidth}
    \captionsetup{position=top,justification=centering}    
    \centering
    \caption{}    
    \includegraphics[width=\textwidth]{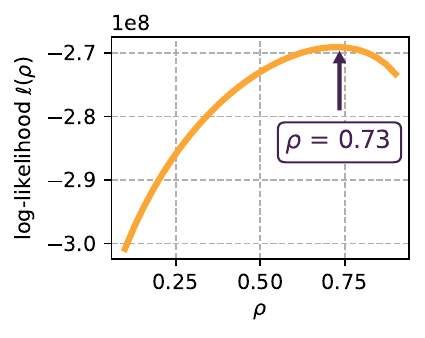}
    \end{subfigure}
    \begin{subfigure}[t]{.31\textwidth}
    \captionsetup{position=top,justification=centering}    
    \centering
    \caption{}    
    \includegraphics[width=\textwidth]{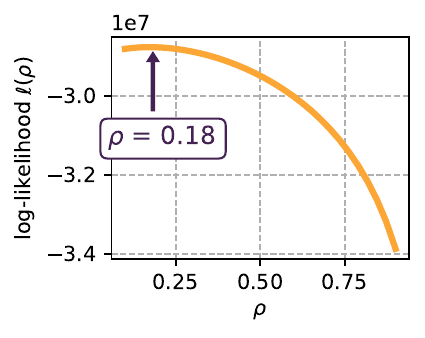}
    \end{subfigure}\\
    \begin{subfigure}[t]{.31\textwidth}
	\captionsetup{position=top,justification=centering}    
	\centering
	\caption{}
	\includegraphics[width=\textwidth]{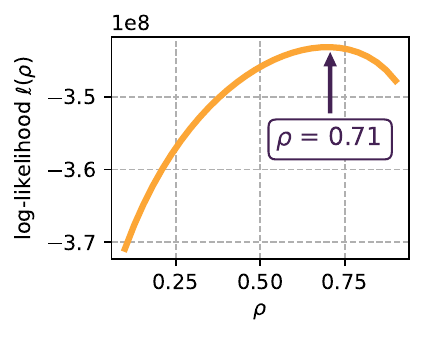}
	\end{subfigure}
	\begin{subfigure}[t]{.31\textwidth}
	\captionsetup{position=top,justification=centering}    
	\centering
	\caption{}    
	\includegraphics[width=\textwidth]{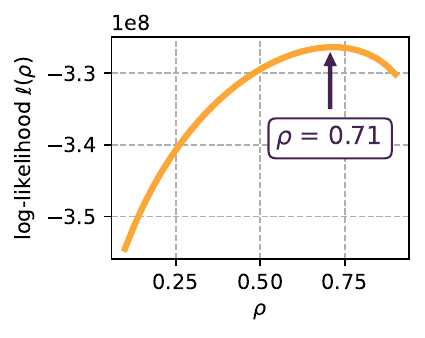}
	\end{subfigure}
	\begin{subfigure}[t]{.31\textwidth}
	\captionsetup{position=top,justification=centering}    
	\centering
	\caption{}    
	\includegraphics[width=\textwidth]{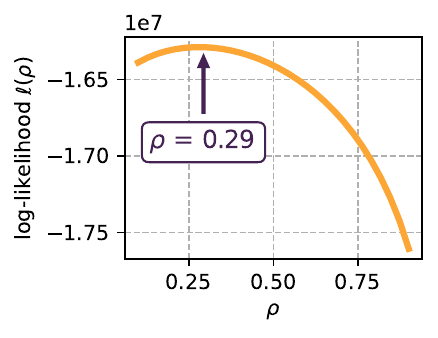}
	\end{subfigure}\\
    \begin{subfigure}[t]{.31\textwidth}
	\captionsetup{position=top,justification=centering}    
	\centering
	\caption{}
	\includegraphics[width=\textwidth]{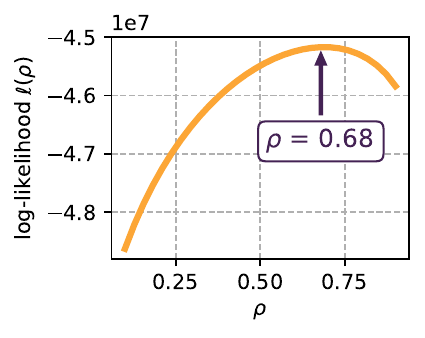}
	\end{subfigure}
	\begin{subfigure}[t]{.31\textwidth}
	\captionsetup{position=top,justification=centering}    
	\centering
	\caption{}    
	\includegraphics[width=\textwidth]{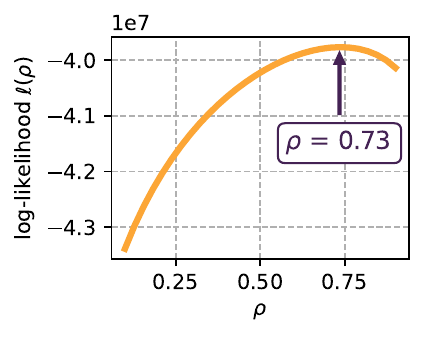}
	\end{subfigure}
	\begin{subfigure}[t]{.31\textwidth}
	\captionsetup{position=top,justification=centering}    
	\centering
	\caption{}    
	\includegraphics[width=\textwidth]{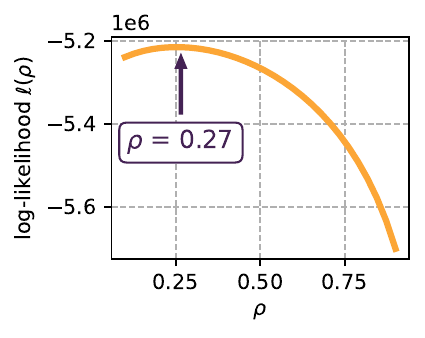}
	\end{subfigure}\\     
    \caption{Log-likelihoods aggregated across all authors for DBLP (top row), Nature (middle row), and PRE (bottom row). Panels (a), (d), and (g) show external and self-citations combined; (b), (e), and (h) show external citations only; (c), (f), and (i) show self-citations only.}\label{fig:all-aggregated}    
\end{figure}
As a first step, we examined citation distributions across all authors in each dataset. For each author, we computed the log-likelihood with Eq.~(\ref{eq:logl}) and then aggregated these to obtain a dataset-level log-likelihood via Eq.~(\ref{eq:loglall}). We estimated $\rho$, the fraction of preferential citations, by maximising this combined log-likelihood. The results are shown in Figure~\ref{fig:all-aggregated}. In panel (a), the log-likelihood was calculated for the combined set of external and self-citations in the DBLP dataset. As evident, there is a strong bias toward preferential attachment, with the maximum log-likelihood corresponding to $\rho \approx 0.68$. This strongly supports the hypothesis that the "rich get richer" mechanism plays a significant role in the dynamics of citation distribution.

An interesting extension of this analysis is to focus exclusively on either external or self-citations. This can be achieved by reconstructing the citation vectors as before - event by event - while restricting the log-likelihood calculations to a single type of citation. Specifically, this involves limiting the innermost summation in Eq.~(\ref{eq:logl}) to one type of citation, while still retaining both types of citations in the citation vector $X$. Panel (b) of Figure~\ref{fig:all-aggregated} shows the results for external citations for DBLP. The maximum log-likelihood shifts further to the right compared to panel (a), corresponding to $\rho \approx 0.73$. This observation leads to two conclusions. First, preferential attachment plays a more pronounced role in the distribution of external citations. Second, the shift in the maximum log-likelihood between panels (a) and (b) suggests that self-citations are associated with smaller values of $\rho$.

This hypothesis is confirmed in panel (c), which displays the log-likelihood calculated exclusively for self-citations in the DBLP dataset. Here, we observe an almost inverse relationship to the trends in the previous panels. The maximum log-likelihood is positioned toward the left of the plot, corresponding to $\rho \approx 0.18$. This indicates that while preferential attachment is a dominant mechanism for external citations, it is far less influential for self-citations. Instead, self-citations appear to follow a different set of distribution rules, distinct from those governing preferential attachment.

The results for Scopus-based datasets are also presented in Figure~\ref{fig:all-aggregated}. As shown, the patterns are qualitatively consistent with those observed in the DBLP dataset. Quantitatively, the proportion of preferential self-citations appears to be higher in the Scopus datasets than in DBLP. We will return to this finding in the last section.

\subsection{Analysis of individual authors}

\begin{figure}[!ht]
    \centering
    \begin{subfigure}[t]{.31\textwidth}
    \captionsetup{position=top,justification=centering}    
    \centering
    \caption{}
    \includegraphics[width=\textwidth]{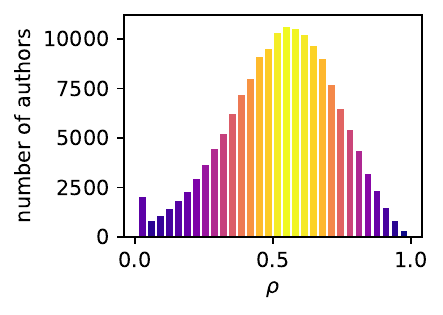}
    \end{subfigure}
    \begin{subfigure}[t]{.31\textwidth}
    \captionsetup{position=top,justification=centering}    
    \centering
    \caption{}    
    \includegraphics[width=\textwidth]{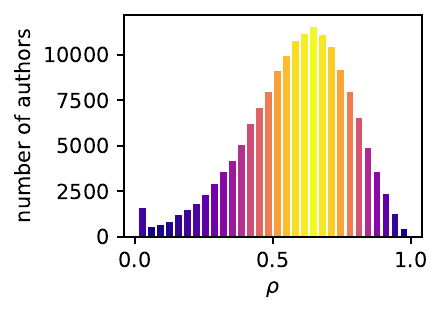}
    \end{subfigure}
    \begin{subfigure}[t]{.31\textwidth}
    \captionsetup{position=top,justification=centering}    
    \centering
    \caption{}    
    \includegraphics[width=\textwidth]{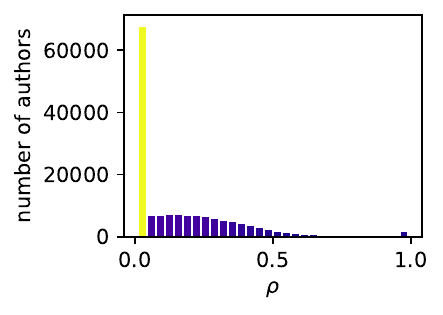}
    \end{subfigure}\\
    \begin{subfigure}[t]{.31\textwidth}
	\captionsetup{position=top,justification=centering}    
	\centering
	\caption{}
	\includegraphics[width=\textwidth]{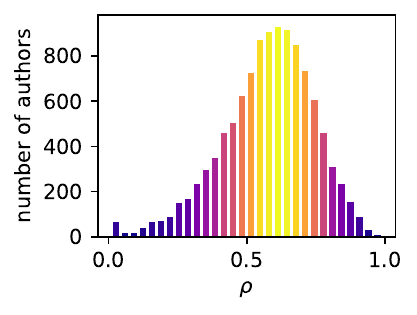}
\end{subfigure}
\begin{subfigure}[t]{.31\textwidth}
	\captionsetup{position=top,justification=centering}    
	\centering
	\caption{}    
	\includegraphics[width=\textwidth]{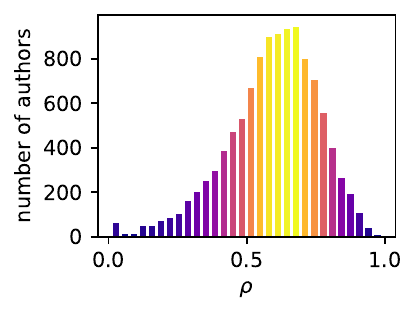}
\end{subfigure}
\begin{subfigure}[t]{.31\textwidth}
	\captionsetup{position=top,justification=centering}    
	\centering
	\caption{}    
	\includegraphics[width=\textwidth]{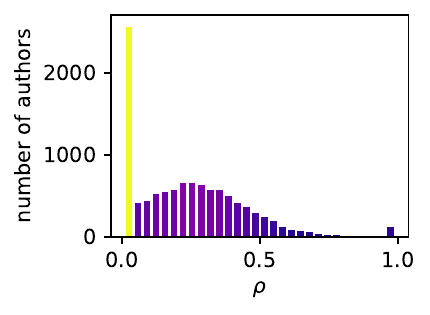}
\end{subfigure}\\
    \begin{subfigure}[t]{.31\textwidth}
	\captionsetup{position=top,justification=centering}    
	\centering
	\caption{}
	\includegraphics[width=\textwidth]{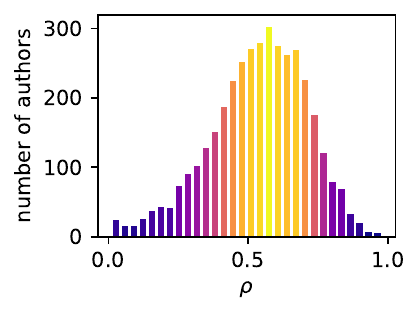}
\end{subfigure}
\begin{subfigure}[t]{.31\textwidth}
	\captionsetup{position=top,justification=centering}    
	\centering
	\caption{}    
	\includegraphics[width=\textwidth]{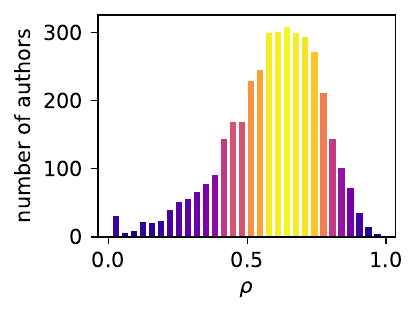}
\end{subfigure}
\begin{subfigure}[t]{.31\textwidth}
	\captionsetup{position=top,justification=centering}    
	\centering
	\caption{}    
	\includegraphics[width=\textwidth]{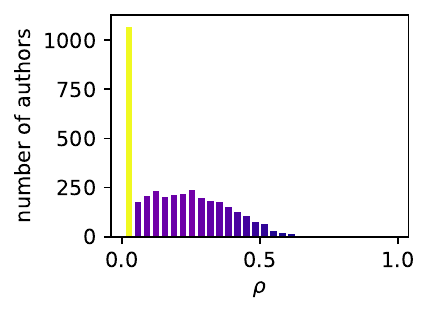}
\end{subfigure}\\        
    \caption{Distributions of $\rho$ for individual authors for DBLP (top row), Nature (middle row), and PRE (bottom row). Panels (a), (d), and (g) show external and self-citations combined; (b), (e), and (h) show external citations only; (c), (f), and (i) show self-citations only.}\label{fig:individual}    
\end{figure}
The analysis of the log-likelihood presented in the previous section can also be performed at the level of individual scientists. This involves processing the citation events for a single scientist, constructing the log-likelihood based on these events using Eq.~(\ref{eq:logl}), and determining the value of $\rho$ that corresponds to the maximal value of the log-likelihood. The results of this procedure are shown in Figure~\ref{fig:individual}, which presents histograms of $\rho$ for all scientists who received more than 50 citations and published at least 10 papers. We added a citation-count threshold, alongside the paper-count threshold, to exclude authors with short records - most would otherwise be classified as non-PAR based on their $\rho$ values.

Panels (a), (b), and (c) of Figure~\ref{fig:individual} displays the distribution of $\rho$ when both external and self-citations are included in the log-likelihood. The distribution is biased toward higher values of $\rho$, supporting the findings from the previous section based on the aggregate log-likelihood for the entire dataset. The mean value of $\rho$ is $0.53$ for DBLP, $0.57$ for Nature, and $0.53$ for PRE. However, a natural question arises: why is there a notable discrepancy between the average $\rho$ calculated for individual scientists and the maximum $\rho$ obtained from optimising the log-likelihood of the entire dataset? The answer lies in the aggregation process. In the previous subsection, the log-likelihood of each citation event contributed to the total value being optimized - often multiple times, as each citation event was processed for every author of the cited article. Consequently, authors with many citations had a larger influence on the aggregate log-likelihood than those with fewer citations. In contrast, the individual-author approach in this subsection treats all authors equally, regardless of their citation counts, leading to the observed differences.

The distribution of $\rho$ for external citations, shown in panels (b), (e), and (h), aligns with the results from the previous subsection. It is skewed toward higher values, with an average $\rho$ of $0.58$ for DBLP, and $0.59$ for both Nature and PRE, indicating that preferential attachment plays a significant role in the distribution of external citations. Conversely, the distribution of $\rho$ for self-citations, depicted in panels (c), (f), and (i), is concentrated around lower values, with an average of $0.17$ for DBLP, $0.25$ for Nature, and $0.20$ for PRE. Notably, a substantial fraction of scientists (approximately 40\%) are characterised by $\rho$ values close to 0 (as shown in the first bin of the histogram). These findings further reinforce the conclusion that self-citations are fundamentally different from external citations, as they do not follow the "rich get richer" dynamic.

\begin{figure}[!ht]
	\centering
	\includegraphics[width=\linewidth]{ 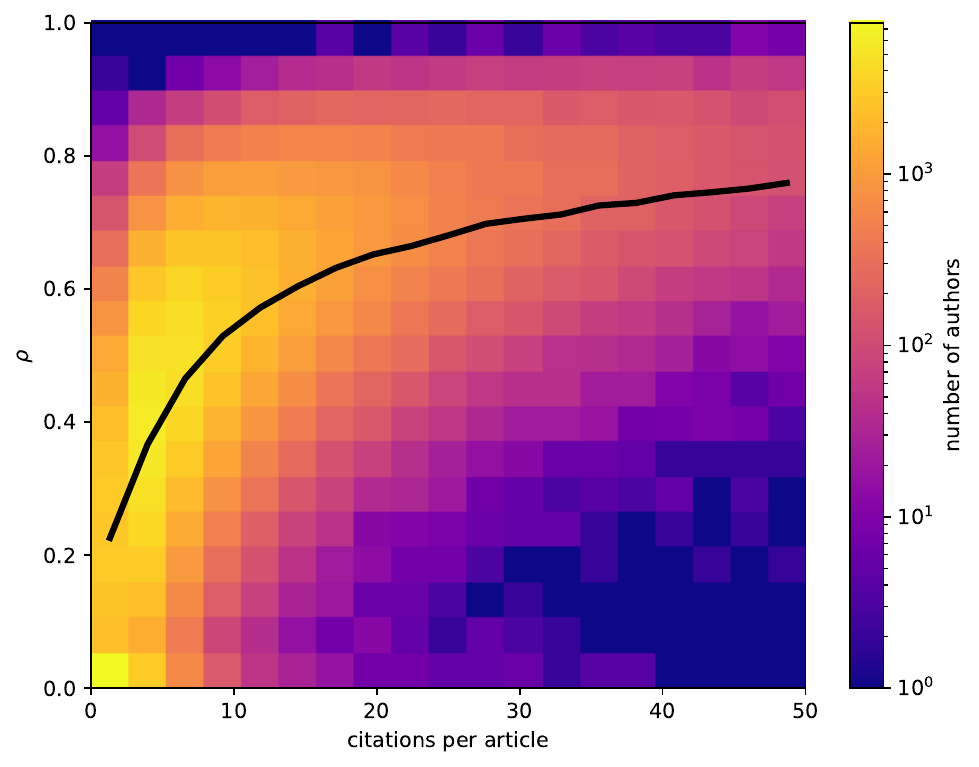}
	\caption{2D histogram of $\rho$ (the probability of preferential citation for both external and self-citations combined) as a function of the number of citations per article for the DBLP dataset, with the black solid line indicating the average $\rho$ for a given number of citations per article.}\label{fig:rhocit}
\end{figure}
Lastly, it is reasonable to assume that different groups of scientists - for instance, those with varying levels of prominence - are characterised by distinct $\rho$ values. However, quantifying prominence introduces challenges. While many bibliometric indicators exist for this purpose \cite{Todeschini2016}, the most well-known being the Hirsch index (or $h$-index; \textcite{Hirsch2005}), such indicators are often controversial. However, they remain important factors in decisions regarding promotions and funding allocation \cite{Demetrescu2020, Bornmann2008, Kelly2006}, although there are initiatives that advocate for moving toward a qualitative and context-rich assessment \cite{torressalinas2024}. That said, calculating bibliometric indices on a limited subset of the citation network may be problematic. For example, this issue is mitigated in studies like \textcite{Siudem2020}, where the number of citations is derived directly from article metadata rather than reconstructed from the network. To avoid potential misinterpretations stemming from inconsistent citation counts, we opted for a straightforward measure of prominence (or, more accurately in this context, popularity or "citability"): the average number of citations per paper. Figure~\ref{fig:rhocit} illustrates how $\rho$ changes, on average, with this measure for the DBLP dataset (the results for the Scopus-based datasets are qualitatively in agreement). As shown, $\rho$ increases with the average number of citations per paper. This suggests that the more citable a scientist is, the more preferential the citation of their papers becomes. This intuitive result will be revisited in the Discussion section.

\section{Discussion}

In this manuscript, we investigated whether the preferential attachment rule (commonly referred to as the "rich get richer" or Matthew effect) is the primary driving force behind the distribution of citations to scientific articles. By examining individual citations one by one, we calculated the probability that a citation is assigned according to PAR. Our analysis was conducted on a large combined set of authors as well as for individual scientists. This was made possible through the DBLP dataset, a standard source of data for such studies, and two Scopus-based datasets.

Our findings indicate that, at least for the aggregated dataset of multiple scientists, the preferential attachment rule is indeed the dominant factor in the citation distribution process, with the probability of preferential citations close to 70\%. This reinforces the foundational assumption behind models such as the Ionescu-Chopard model, which often incorporate some form of PAR. However, when focusing on individual scientists rather than the combined dataset, a more nuanced picture emerges. Specifically, we observe a spectrum where some scientists attract significantly more preferential citations than others, particularly with respect to external citations. Nonetheless, the average probability of a preferential external citation remains high (when compared to self-citations).

When the analysis is restricted to self-citations, a noticeably different behaviour becomes apparent. For the aggregated dataset, the probability of a preferential self-citation is approximately 20\% - 30\%. The difference between external and self-citations becomes even more pronounced at the individual level. A substantial group of authors exhibits a probability of preferential self-citation close to 0, and only rarely does this probability exceed 50\%. These findings strongly suggest that self-citations represent a distinct category of citations, governed by rules that differ from those of external citations.

Interestingly, the prominence of the preferential attachment rule appears to increase with an author's citability, defined here as the average number of citations per paper. It seems that more citable authors are cited in a more preferential manner, while the citation patterns for less-cited authors tend to involve greater randomness. This result is intuitive: highly citable authors likely produce influential work, some of which frequently appears in bibliographies. Scientists who encounter these papers repeatedly are naturally inclined to cite them in their own manuscripts, thereby amplifying the Matthew effect. Additionally, articles with higher citation counts are more likely to be recommended by search engines. Conversely, less-cited authors producing important but specialised work are more likely to experience citation patterns that appear less preferential and more uniform, as their work is cited primarily within specific contexts. Finally, the expectation to include a sufficient number of references in bibliographies may lead to the occasional inclusion of less-relevant citations \cite{Herrera2021}, which could partially explain the randomness in the citation patterns of less-citable authors \cite{Simkin2007citing}.

However, self-citations once again diverge from this pattern. When we repeated the analysis shown in panel (d) of Figure~\ref{fig:individual} for self-citations, no meaningful correlation similar to that observed for the full citation set or external citations could be detected. This suggests that, irrespective of an author's citability, the preferential attachment rule does not explain the distribution of self-citations.

\begin{figure}[!ht]
    \centering
    \includegraphics[width=\linewidth]{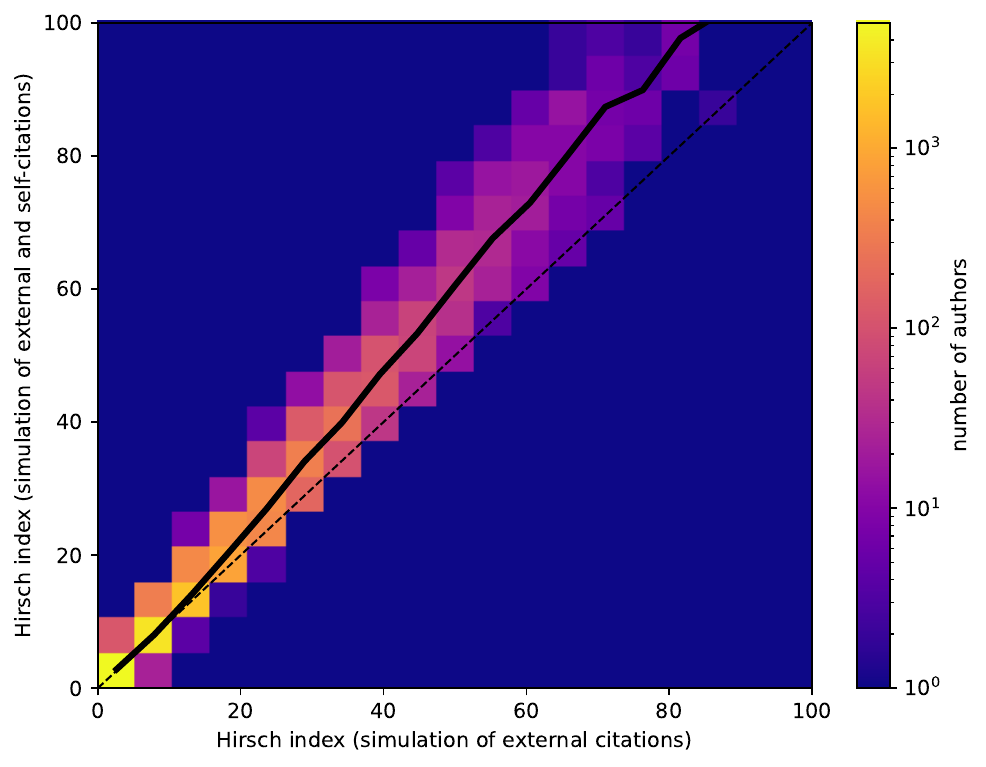}
    \caption{Results of citation distribution simulations. The $Y$-axis represents the $h$-index calculated after distributing both external and self-citations, while the $X$-axis represents the $h$-index calculated using only external citations. Colours indicate a histogram of occurrences for each $h$-index pair, with the black solid line showing the average. The dashed line corresponds to $Y = X$. For aesthetic purposes, bins with a count of 0 are displayed in the same colour as those with a count of 1.}\label{fig:sim}
\end{figure}
Finally, we would like to address an important issue related to self-citations. Our results clearly demonstrate that self-citations represent a distinct category of citations, governed by rules different from those of external citations. However, how significant are they in the broader context? Scientists can use self-citations to enhance certain bibliometric measures - potentially, though not necessarily, in a problematic manner. One could argue that self-citations could simply be excluded from citation vectors before calculating these measures. To assess the potential impact of self-citations, we conducted additional simulations using the Scopus - Nature dataset.

For each scientist, we reconstructed (or rather, in this case, simulated) their citation vector by processing incoming citations sequentially, year by year, in the same manner used to estimate $\rho$. However, instead of assigning citations to actual cited papers based on real data, we distributed citations according to predefined rules. We assumed that external citations were always distributed according to the preferential attachment rule - an extreme assumption but justified by the results presented in this paper. In contrast, self-citations were distributed uniformly across the scientist's previously published papers.

We performed two types of simulations. In the first, we omitted self-citations entirely and distributed only external citations. In the second, we distributed both external and self-citations but removed the self-citations from the final simulated citation vector. At the conclusion of each simulation, we calculated the Hirsch index ($h$-index) for each scientist, effectively using only external citations (as self-citations were either not distributed or subtracted from the citation vector). Figure~\ref{fig:sim} presents the results, where the $Y$-axis corresponds to the $h$-index calculated after distributing both external and self-citations, and the $X$-axis corresponds to the $h$-index calculated using only external citations.

Two important conclusions can be drawn from this figure. First, the Hirsch index is, on average, higher when self-citations are included, in some extreme cases by nearly 50\%. This suggests that self-citations can significantly enhance the visibility and perceived scientific impact of a researcher’s work. While this process can occur naturally - as in our simulations, where a benign distribution scheme was assumed - it is easy to imagine scenarios where self-citations are strategically manipulated to target specific bibliometric measures.

The second conclusion, which provides valuable insight for regulators, is that self-citations cannot simply be removed from citation vectors. While they can be excluded arithmetically, doing so merely removes their numerical contribution and does not account for the potential influence self-citations have on the distribution of external citations. The existence and distribution of self-citations may shape the patterns of external citations in ways that are both indirect and significant, and this influence cannot be ignored when assessing bibliometric indices.

Looking ahead, many questions about the nature of self-citations remain open. Although the qualitative patterns are consistent, we observe a quantitative discrepancy in self-citation PAR rates between the DBLP dataset and the two Scopus-based datasets - Nature and Physical Review E. Several explanations are possible. One is that DBLP represents an incomplete portion of the full citation network, and the observed differences stem from missing publications and references. Another consideration is the disciplinary scope of each dataset: DBLP primarily covers computer science, Nature is interdisciplinary, and PRE, while also interdisciplinary, predominantly serves the physics community. Given that citation practices vary across disciplines, it is reasonable to expect that the parameter $\rho$ may likewise differ. Testing this hypothesis would be a compelling direction for future research but would require access to a significantly larger and more comprehensive dataset.

Another important consideration is that the results presented in this study are based on aggregated data across all scientists in the datasets - that is, not only researchers from different disciplines, but also different age groups and career stages. It is reasonable to expect, however, that the parameter $\rho$ may vary across these groups. Preliminary analyses of the DBLP dataset support this intuition, showing that $\rho$ differs among cohorts \cite{fronczak2007} defined by the year researchers began their careers: broadly speaking, it increases over time for self-citations and decreases for external citations. Properly analysing such cohorts is a non-trivial task that requires careful methodological treatment and represents a promising direction for future research.

Also, while we have shown that the distribution of self-citations is not governed by preferential attachment, the exact mechanisms by which they are assigned to articles remain unclear. Scientists are known to favour their own work when compiling bibliographies for their papers (the so-called self-citation bias; see \textcite{Brysbaert2011}), and self-citations can account for a significant fraction of total citations. As noted earlier, this makes them a potential tool for artificially inflating bibliometric indices such as the $h$-index \cite{Amjad2020, Gianoli2009}. Developing models in which authors aim to maximise specific bibliometric indicators could be a promising avenue for understanding the distribution of self-citations and identifying authors who engage in such practices. Moreover, research indicates that most self-citations occur shortly after publication and that their influence wanes more rapidly than external citations \cite{Shah2015, Costas2010, Wolfgang2004, Aksnes2003, Lyon1982}. Incorporating a time-dependent component, such as manuscript ageing and history \cite{pan2018}, into our model could enhance its ability to capture these temporal dynamics.  There are also other factors that one might want to consider, such as reputation \cite{petersen2014}. Additionally, exploring whether the asymmetry of interactions \cite{fronczak2022weak, mrowinski2024tie, orzechowski2023} - both within scientific collaboration networks and citation networks - can be linked to and explain the citation behaviours of authors would be an intriguing direction for future research.

\section{Data and code availability}

The source code and the anonymised Scopus-based datasets can be found in \textcite{mrowinski2025citescode}. The 12th version of the DBLP Citation Network Dataset \cite{Tang2008} is publicly and freely available at https://www.aminer.cn.

\printbibliography

\end{document}